# Mirror-Coupled Microsphere can narrow the Angular distribution of Photoluminescence from WS$_2$ Monolayers

Shailendra K. Chaubey,[1] Sunny Tiwari,[1,2] Gokul M. A.,[1] Diptabrata Paul,[1] Atikur Rahman,[1] and G.V. Pavan Kumar[1]

[1]*Department of Physics, Indian Institute of Science Education and Research, Pune-411008, India*

[2]*Current Address -Aix Marseille Univ, CNRS, Centrale Marseille, Institut Fresnel, AMUTech, 13013, Marseille, France*

(*Electronic mail: pavan@iiserpune.ac.in)

(Dated: 8 June 2022)

Engineering optical emission from two-dimensional, transition metal dichalcogenides (TMDs) such as Tungsten disulfide (WS$_2$) has implications in creating and understanding nanophotonic sources. One of the challenges in controlling the optical emission from two-dimensional materials is to achieve narrow angular spread using simple photonic geometry. In this article, we study how the photoluminescence of a monolayer WS$_2$ can be controlled when coupled to film coupled microsphere dielctric antenna. Specifically, by employing Fourier plane microscopy and spectroscopic techniques, we quantify the wavevector distribution in the momentum space. As a result, we show the beaming of the WS$_2$ photoluminescence with angular divergence as low as $\theta_{1/2} = 4.6°$. Furthermore, the experimental measurements have been supported by three-dimensional numerical simulations. We envisage that the discussed results can be generalized to a variety of two-dimensional materials, and can be harnessed for on-chip nonlinear and quantum technology.

Two-dimensional transition metal dichalcogenides (TMDs) have attracted major attention in recent years because of their unique optical and electronic properties[1–3]. High exciton binding energy and direct bandgap make them suitable candidates for optoelectronic device applications[3–5]. TMDs show other unique properties such as valley polarization which make them suitable for valleytronics and spin-orbit interaction studies[6,7].

Influencing and controlling the emission properties of the TMDs are key to improving the efficiency of optoelectronic devices. Recent advancements in the field of plasmonic and dielectric nanoantennas have enabled control over the fundamental emission properties such as wavevector distribution and Purcell enhancement[8–11]. Furthermore, plasmonic nanostructure coupled to TMDs has been studied for strong coupling[12–14], photoluminescence (PL) enhancement[15–18], surface-enhanced Raman scattering[19,20], spectrum tailoring[21,22], trion enhancement,[23] and directional emission[24]. Although plasmonic nanostructure provides great control over the emission properties, they suffer large Ohmic losses and have broad resonances[25]. On the other hand, dielectric structures such as microdisk, cylinder, and microsphere show low absorption and sharp resonances, which make them a suitable candidate for controlling the emission properties without much loss. To this end, studying the optical transition characteristics of TMDs coupled whispering gallery mode of the dielectric microsphere and disk has gained relevance.

A variety of prospects like PL enhancement[26], lasing[26–28], directional emission[29], and out-of-plane dipole excitation[30] have been achieved using such geometries. Recently hybrid metallo-dielectric cavity has been used to enhance light-matter interaction[31–34], build hybrid metasurfaces[35], tailor the quality factor[36], and in designing directional sources[37–40]. Especially, dielectric particle on the mirror is one such geometry which is widely used for hybridizing surface plasmon and whispering gallery modes (WGMs)[35], surface-enhanced Raman scattering[41], studying wavevector and polarization state of emission from fluorescent dye[42], and probing magnetic resonances[43]. Although, basic emission properties of TMDs such as Purcell enhancement and polarization of emission have been studied in great detail but the wavevector distribution is relatively less explored using geometries involving complicated fabrication and etching techniques[44].

In this context, tunability of directionality of the optical emission from nanostructures coupled to TMDs remains a challenging problem, and highly directional emission from TMDs is yet to be achieved. Motivated by this, we study the optical emission properties of a monolayer WS$_2$ coupled to a metallo-dielectric antenna formed by placing a dielectric microsphere on a gold mirror. Specifically, we report the PL beaming from monolayer WS$_2$ coupled to this antenna. We study the wavevector distribution of monolayer WS$_2$ coupled to a metallo-dielectric antenna. Using Fourier plane imaging, we characterized the PL beaming and found the divergence angle to be as narrow as $\theta_{1/2} = 4.6°$. Furthermore, we numerically study the near-field and far-field properties by using finite element method (FEM) based numerical simulations.

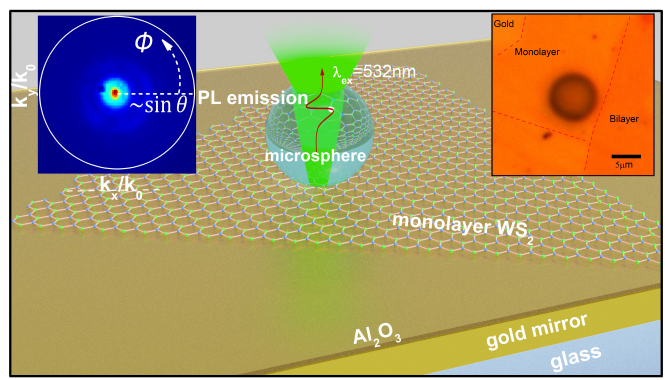

FIG. 1. Schematic of the experimental configuration. A WS$_2$ monolayer was placed over a gold mirror separated by a 3 nm Al$_2$O$_3$ spacer layer, over which polystyrene microspheres of diameter 7 μm were dropcasted. A single microsphere was excited using a 532 nm laser. Insets show the bright field (right) and Fourier plane (left) images. Boundaries of the monolayer are shown with red dotted lines.

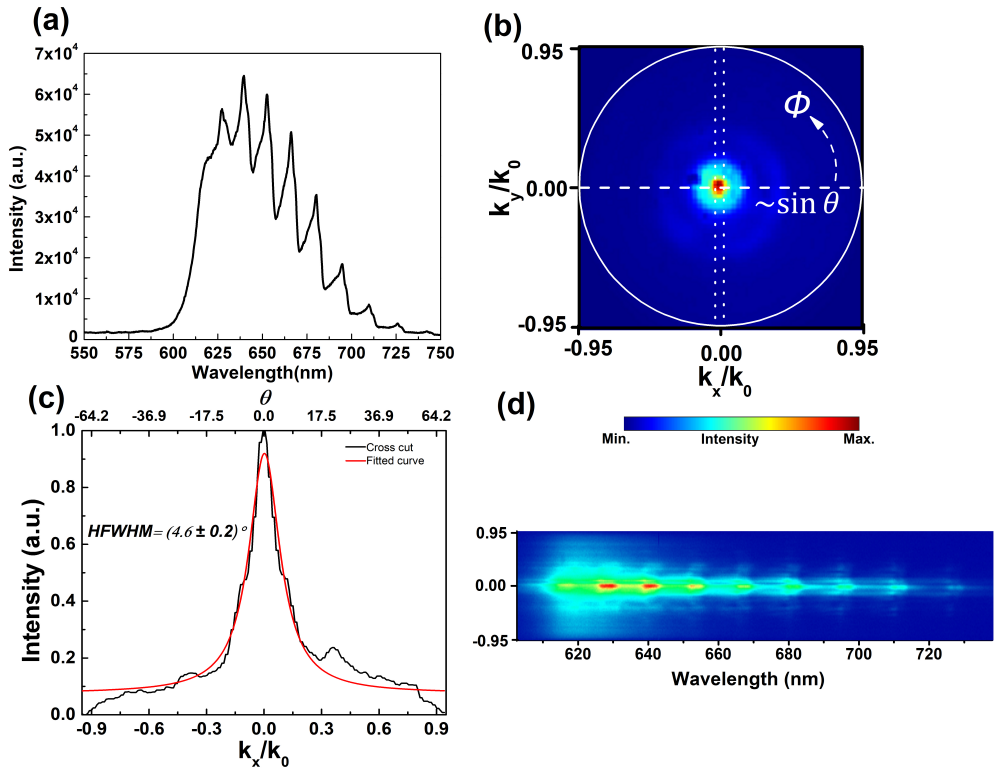

FIG. 2. Beaming of WS$_2$ PL. (a) PL spectrum of WS$_2$ monolayer coupled to a metallo-dielectric antenna. (b) Fourier plane image of the PL emission from the WS$_2$ coupled to antenna. (c) Intensity profile of the Fourier plane image along the white dashed horizontal line in the Fourier plane image. (d) Energy-momentum image of the PL emission along the white vertical rectangular region.

Figure 1 shows the schematic of the experimental configuration. Gold mirrors were prepared by thermal vapor deposition (TVD) by depositing 160 nm gold film on a glass coverslip. Monolayer WS$_2$ were grown using chemical vapor deposition (CVD)[45,46] and characterized using Raman and PL spectroscopy (See Supplementary Information S1). Raman and PL spectrum match well with the reported Raman and PL spectra[47,48]. WS$_2$ flakes were transferred to Al$_2$O$_3$ coated gold mirror using a wet chemical method[49]. Briefly, polystyrene solution is spin-coated on the WS$_2$ sample followed by baking at 100° C. Sample was placed in water, and WS$_2$ was fished on the desired substrate with polystyrene support film when it started floating in the water. Polystyrene was removed using toluene, followed by cleaning the sample using acetone and IPA. A 4 nm Al$_2$O$_3$ layer is used between WS$_2$ and the mirror to avoid the charge screening, which might lead to PL quenching. Al$_2$O$_3$ spacer layer was deposited using atomic layer deposition. Polystyrene microspheres purchased from Sigma-Aldrich were dropcasted on a monolayer WS$_2$ placed on a gold mirror. A single microsphere placed on the mirror was excited by a 532 nm laser using a 100x, 0.95 numerical aperture (NA) objective lens. The backscattered signal was collected using the same objective lens and projected onto EMCCD and spectrometer for imaging and spectroscopy. Microsphere placed on a gold mirror acts as a directional metallo-dielectric antenna to direct the WS$_2$ PL to a narrow range of wavevectors which was probed using Fourier plane imaging[50,51]. A combination of edge and notch filters were used in the output path to efficiently reject the elastically scattered light, which can be observed in the PL spectrum shown in supplementary information S1. Details of the experimental setup can be found in reference[42].

Figure 2(a) shows the PL spectrum collected from the WS$_2$ coupled to the metallo-dielectric antenna. Microsphere was excited using a 532 nm tightly focused laser beam. The microsphere focuses the incoming laser to a small region because of the photonic nanojet effect[52–54], which excites the PL of the WS$_2$ monolayer. The PL emission from the monolayer acts as a near-field source which excites the WGMs of the microsphere. WGMs are the sharp resonances supported by the dielectric microstructures[28]. WGMs of the microsphere can be seen as sharp peaks riding over a broad PL spectrum.

To study the wavevector distribution of the PL emission from the metallo-dielectric antenna, Fourier plane microscopy and spectroscopy were performed. Fourier plane image maps the wavevectors of the light in terms of $\theta$ and $\phi$ coordinates. The radial coordinate in the Fourier plane image is the numeri-cal aperture (NA)= n sin($\theta$), and $\phi$ is the azimuthal coordinate that varies from 0 to $2\pi$. Figure 2 (b) shows the Fourier plane image of the PL emission. It can be observed from figure 2 (b) that the $\theta$ spread is very small, and the majority of the emission is beaming at the center of the Fourier plane image.

The PL generated below the microsphere gets out-coupled from the upper side of the microsphere in a narrow range of angles because of the reverse of the photonic nanojet effect[53]. To quantify the spread of the emission wavevectors, we plot the intensity profile along the white dotted line in the Fourier plane image (Figure 2(c)). The measured intensity profile was fitted using the Lorentzian function, which shows that the divergence angle, $\theta_{1/2}$, half of the full width at half maxima (H-FWHM), of the emission is only 4.6°. For various microspheres (five microspheres) the divergence angle varies slightly with an average of $(5.2 \pm 0.6)°$. (See supplementary information S2)

To further confirm that both WGMs coupled and uncoupled emission get directed to a narrow wavevector range from the metallo-dielectric antenna, we performed energy-momentum imaging (figure 2(d)). For this, the Fourier plane image (figure 2(b)) was projected onto the slit of the spectrometer to get the wavelength information[55–57]. The energy-momentum image shows that both WGMs coupled and uncoupled emission is out-coupling from the microsphere towards the center of the Fourier plane image with a narrow range of angles. Additionally, this further rejects the possibility of elastically scattered light at the center of the Fourier plane image.

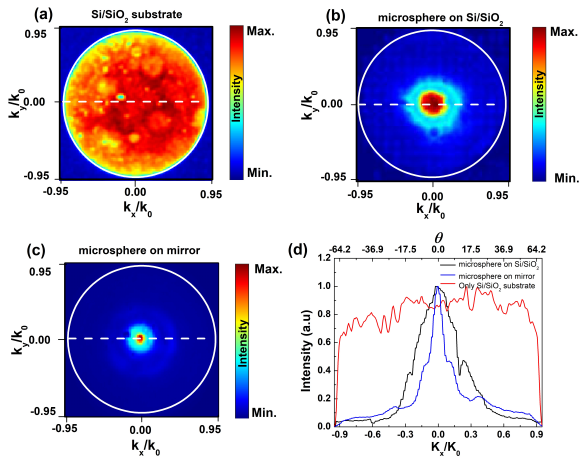

FIG. 3. Significance of using the gold substrate. Fourier plane images (a) on Si/SiO$_2$ substrate, (b) with microsphere on Si/SiO$_2$ substrate, and (c) microsphere on mirror. (d) Intensity profile across the white dotted lines in (a-c).

To highlight the significance of the gold mirror, similar experiments were performed in different configurations. Figure 3(a) shows the wavevector distribution of WS$_2$ PL on the Si/SiO$_2$ substrate. The PL emission from WS$_2$ monolayer is isotropic in nature and covers a broad range of wavevectors. Figure 3(b) shows the wavevector distribution when a microsphere is placed over Si/SiO$_2$ substrate. Placing the monolayer between the microsphere and Si/SiO$_2$ substrate results

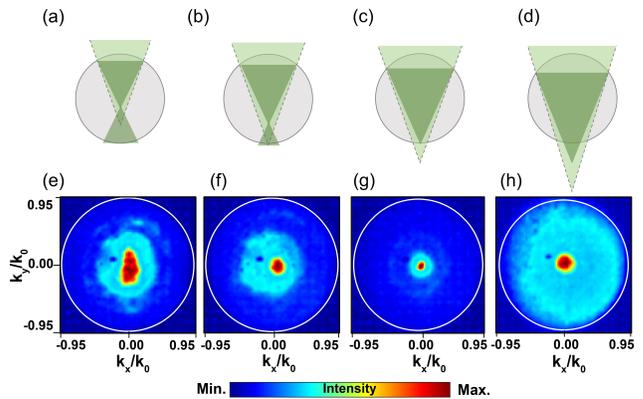

FIG. 4. Effect of focusing of excitation laser in the different planes on angular diversion of emission from the metallo-dielectric antenna. (a-d) Schematic of different focusing schemes and (e-h) are corresponding Fourier plane images. The light green cone represents the focused laser excitation using the objective lens, and the dark green cone shows the effective focusing of the incoming laser by the microsphere.

in confining the wavevectors.

Using a metallic mirror as a substrate further reduces the angular divergence and results in the beaming of emission from the monolayer sandwiched between the microsphere and the mirror (figure 3(c)). This can be understood by the localization of the electric field in the gap between the microsphere and the gold mirror. To understand and corroborate the results, we performed finite element method (FEM) based numerical calculations, which are discussed later in the paper.

Intensity profiles along the white dotted lines in figures 3(a-c) are plotted in figure 3(d). Angular divergence ($\theta_{1/2}$) for the gold and the Si/SiO$_2$ substrate is 4.6° and 11.5°, respectively (See supplementary information S3 for intensity profile fitting). For various microspheres, the divergence angle varies slightly with an average $(13.2 \pm 1.1)°$ for Si/SiO$_2$ substrate (See supplementary information S4). To study the effect of microsphere size on the angular divergence of the emission, we performed experiments on microspheres of sizes 3, 5, 7, and 10 µm. The angular divergence decreased when the microsphere size was chosen to be 7 µm as compared to the divergence obtained with a microsphere of size 3 or 5 µm. A further increase in the microsphere size (10 µm) leads to an increment in the divergence angle. The increment in the divergence angle for the 10 µm microsphere can be understood by an increase in the focal length of the microsphere[52,58]. See supplementary information S5 for details on the variation of angular spreading with a change in the microsphere size.

In photonic nanojet effect, the emission wavevectors are sensitive to the position of the nanojet[58–60]. Focusing dependent experiments were performed to probe the wavevectors as a function of the beam waist of the focused light through the microsphere. Input light was focused in the different planes of the microsphere, and wavevector distribution of PL emission was studied. Figure 4 (a-d) are the different focusing schemes, and (e-h) are the corresponding Fourier plane images. The light green cone represents the focusing of the exci-tation laser beam using the objective lens, while the dark green cone shows the effective focusing of laser beam. It is observed that the divergence was smallest when the input laser was focused slightly below the microsphere. The smallest angular divergence for the position of the waist just outside the microsphere confirms that the beaming of the emission is caused by the reverse of the photonic nanojet effect of the microsphere. An increase in the angular divergence upon defocusing can be understood from the three dipole model given in Supplementary information S6. To mimic the focused and defocused excitation three dipoles were placed at the center of the microsphere separated by 80 nm and 240 nm from each other respectively. The angular divergence is relatively less, when the distance between dipoles increases. The waist size of the excitation Gaussian beam is another parameter which affects the photonic nanojet[58,59]. To probe it experimentally, NA dependent experiment is performed. The microsphere was excited using three different objectives with magnification and NA 50x, 0.5, 100x, 0.8, and 100x, 0.95, however no significant changes are observed (See supplementary information S7).

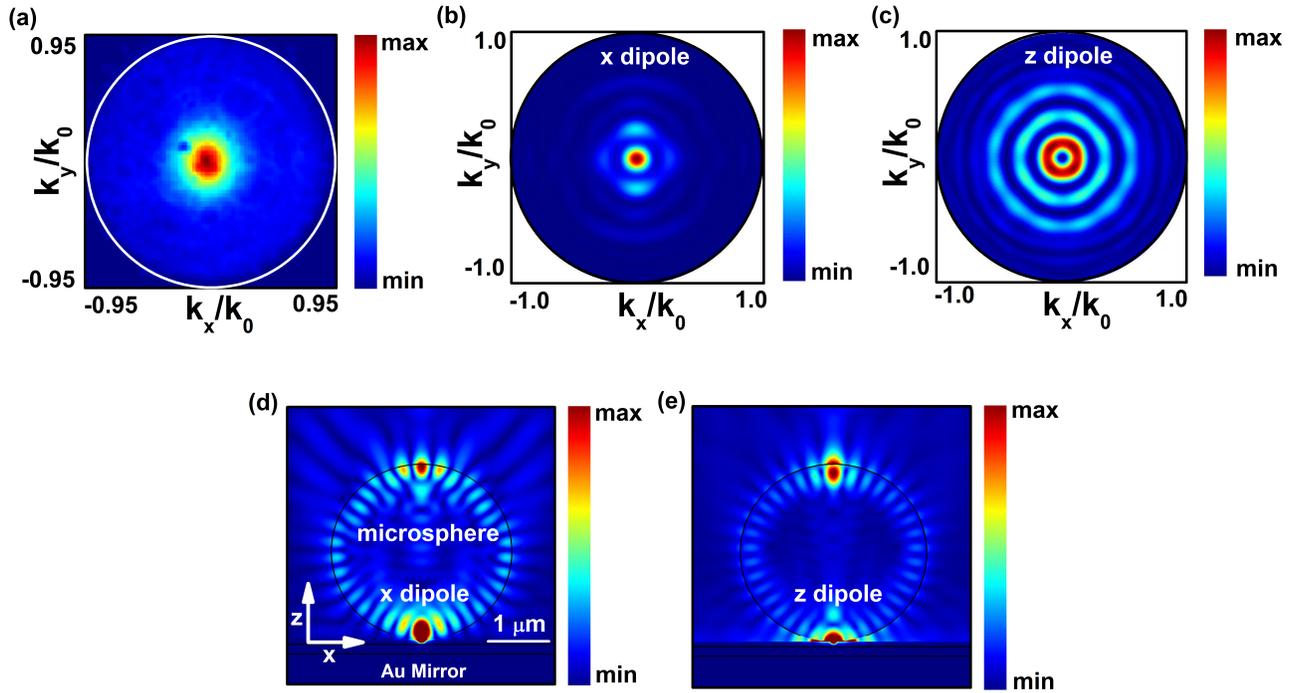

FIG. 5. Calculated Fourier plane imaging of emission from dipoles placed between microsphere and mirror. (a) Experimentally measured Fourier plane image of emission from a TMD monolayer sandwiched between a 3 µm polystyrene microsphere and 160 nm gold mirror. Calculated Fourier plane image of emission from a dipole oscillating in-plane (x-axis) (b) and out-of-plane (z-axis) (c) Calculated electric near-field of a 3 µm polystyrene microsphere placed on a gold mirror excited by placing an oscillating dipole at 645 nm wavelength with in-plane (x-axis) (d) and out-of-plane (z-axis) dipole in the gap between microsphere and mirror (e).

To corroborate and understand the experimental findings, we performed finite element method based numerical calculations in COMSOL Multiphysics. We performed near-field calculations in the COMSOL and calculated the Fourier plane images by using reciprocity arguments[61]. We calculated the near-field profile of a 3 µm diameter polystyrene microsphere placed on a 160 nm thick gold mirror. The gap between the microsphere and gold mirror was set as 5 nm for accounting for the $Al_2O_3$ and molecular layers. Scattering boundary condition was used in the simulations to get rid of the back reflection. For the calculations of electric near-field in wave excitation geometry, the 'extremely fine meshing' configuration was used in the COMSOL. User controlled meshing with a minimum mesh size of 3 nm in the gaps between the microsphere and gold mirror was introduced when calculating the electric near-field in the case of dipole excitation. The maximum mesh size was truncated with $\lambda/20$.

Figure 5(a) shows the experimentally measured Fourier plane image of the emission from a $WS_2$ monolayer coupled to a metallo-dielectric antenna formed by a 3 µm polystyrene microsphere and gold mirror. The electric near-field of a 3 µm polystyrene placed on a gold mirror was calculated upon excitation by placing in-plane and out-of-plane dipoles. We placed an oscillating dipole at 645 nm oriented along the in-plane (x-axis) (figure 5(d)) and out-of-plane (z-axis) (figure 5(e)). The wavelength of oscillation was chosen to be 645 nm, as the PL emission has a peak around this wavelength. The experimental Fourier plane image figure 5(a) matches well with the calculated Fourier plane image for the in-plane dipole figure

5(b). The electric field is more confined when the substrate is a gold mirror as compared to a glass substrate. This is confirmed by performing numerical calculations by exciting microsphere placed on glass and gold substrates with a focused Gaussian beam (See supplementary information S8).

To see how the localization of electric near-field leads to the narrowing of wavevectors, three dipoles were placed at the center of the microsphere separated by 80 nm from each other in the first case and 240 nm in the second case. Calculated electric near-field was projected to the far-field by using reciprocity arguments. The spreading of the wavevectors is less in the former case as compared to the spreading obtained in the latter. (See Supplementary Information S6). This explains the narrowing down of the wavevectors in the case of microsphere placed on a gold mirror.

For in-plane dipole (x-axis), the calculated Fourier plane image matches very well with the experimentally obtained Fourier plane image. The majority of the light is coming out in a beaming manner and is confined to a very narrow range of wavevectors. The calculated Fourier plane image for the out-of-plane dipole shows that the emission is coming out in a doughnut shaped pattern and is also directed towards the center of the Fourier plane image. The results show that the contribution of in-plane dipoles is more pronounced in the case of metallo-dielectric antenna.

To summarize, we have experimentally studied the wavevector distribution of PL coupled to a metallo-dielectric antenna. We showed that the PL emission of $WS_2$, sandwiched between microsphere and gold mirror out-couples in a beaming manner. Microsphere on the mirror geometry acts as an excellent metallo-dielectric antenna to direct the emission from TMDs in a very small range of wavevectors. In the Fourier plane image, the angular divergence $\theta_{1/2} = 4.6°$ was achieved. The experimental findings were corroborated with three-dimensional finite element method based numerical simulations. The proposed hybrid directional antenna will find relevance in two-dimensional material based nanophotonic chips and nano-lasers applications.

## I. SUPPLEMENTARY MATERIAL

See supplementary material for additional information on field calculation, optical characterisation, size variation, multiple measurements and NA dependent experiment.

## ACKNOWLEDGMENTS

SKC and GVPK would like to thank Mandar Deshmukh and Mahesh Gokhale from TIFR, Mumbai for atomic layer deposition of $Al_2O_3$. Authors also thank Chetna Taneja and Vandana Sharma for fruitful discussion. This work was partially funded by DST Energy Science grant (SR/NM/TP-13/2016), Air Force Research Laboratory grant (FA2386-18-1-4118 R and D 18IOA118) and Swarnajayanti fellowship grant (DST/SJF/PSA02/2017-18) to GVPK. AR acknowledges funding support from the Indo-French Centre for the Promotion of Advanced Research (CEFIPRA), project no. 6104-2.

## DATA AVAILABILITY STATEMENT

The data that support the findings of this study are available within the article and its supplementary material

# Supplementary Information

## Mirror-Coupled Microsphere can narrow the Angular distribution of Photoluminescence from WS$_2$ Monolayers

*Shailendra K. Chaubey\*, Sunny Tiwari[1], Gokul M. A., Diptabrata Paul, Atikur Rahman, G.V. Pavan Kumar* *

Department of Physics, Indian Institute of Science Education and Research, Pune-411008, India

[1]Current Address -Aix Marseille Univ, CNRS, Centrale Marseille, Institut Fresnel, AMUTech, 13013, Marseille, France

\*E-mail: shailendrakumar.chaubey@students.iiserpune.ac.in

\*E-mail: pavan@iiserpune.ac.in

**Table of content:**

**S1. Optical characterization of monolayer WS$_2$**

**S2. Reproducibility of the PL beaming from multiple samples**

**S3. Significance of gold mirror as a substrate**

**S4. Multiple Fourier plane images for microsphere on Si/SiO$_2$**

**S5. Microsphere size dependence of angular divergence**

**S6. Electric near-field distribution and far-field Fourier plane image**

**S7. Variation of angular divergence with the numerical aperture of the objective lens**

**S8. Electric near-field distribution on glass and gold mirror after focusing the laser beam through microsphere**

## S1. Optical characterization of monolayer $WS_2$

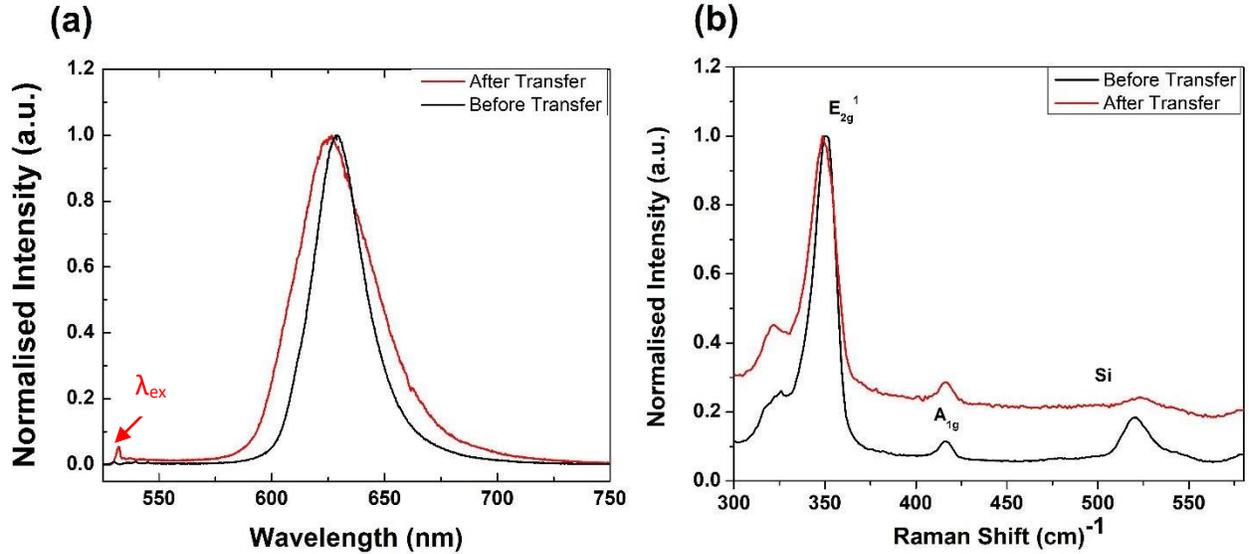

**Figure S1**. Optical characterization of $WS_2$ (a) Photoluminescence (PL) spectrum of $WS_2$ before and after the transfer to the gold mirror. The arrow represents the excitation wavelength, (b) Raman spectrum of the $WS_2$ before and after the transfer. A large PL intensity confirms that $WS_2$ probed is a monolayer. After transferring the monolayer onto the gold mirror, the emission becomes slightly broader while the Raman spectrum remains the same. In the Raman spectrum, Peak $E_{2g}^1$ and $A_{1g}$ are positioned at 351 cm$^{-1}$ and 417 cm$^{-1}$, respectively, and the intensity of the $E_{2g}^1$ is an order of magnitude higher than $A_{1g}$, which matches well with reported Raman spectra [1,2]. In the Raman spectrum of $WS_2$ after the transfer, there is a small peak near 520 cm$^{-1}$, which might be because of the small remains of Si after transfer.

## S2. Reproducibility of the PL beaming from multiple samples:

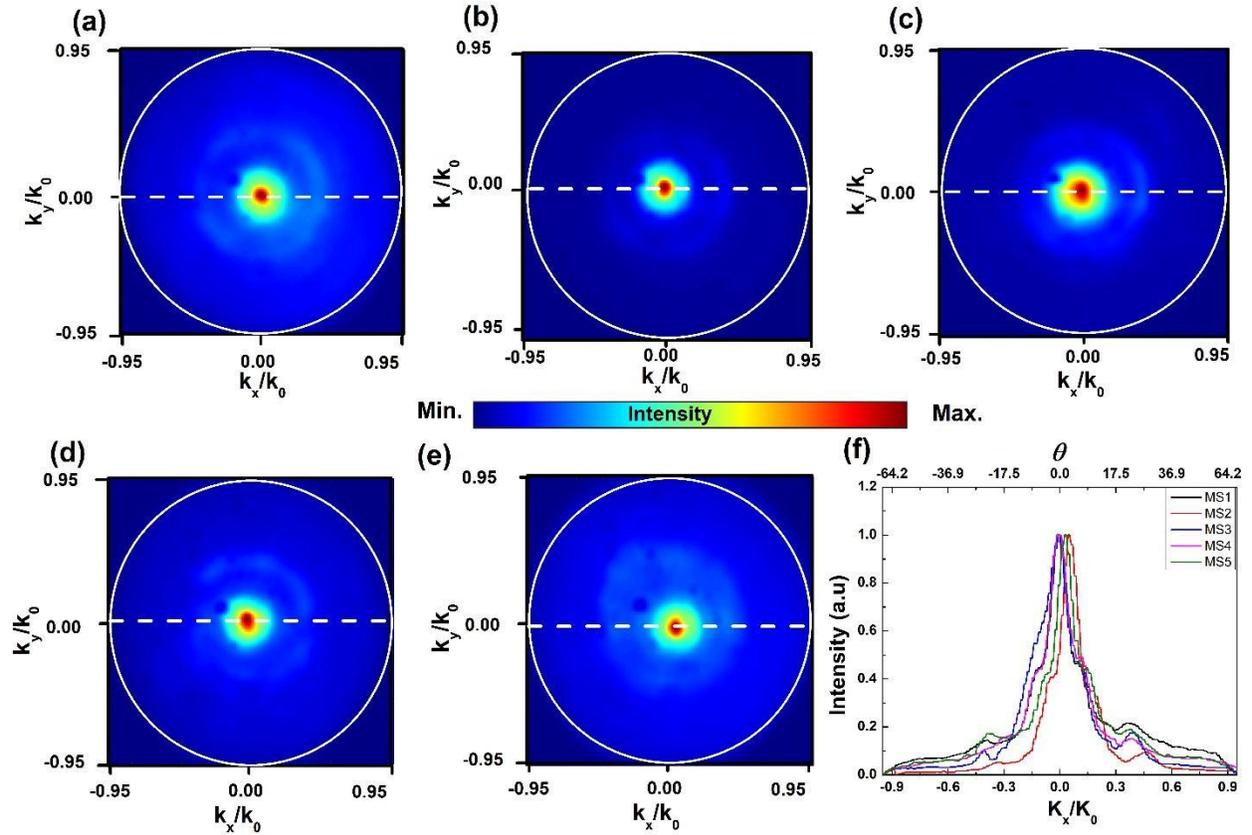

**Figure S2.** Multiple measurements showing the beaming of the emission, (a-e) Fourier plane image showing the PL beaming from five different microspheres. (f) Shows the intensity profile along the white dotted line. The divergence angle was calculated by fitting the Lorentzian function for each intensity profile. The average value of divergence angle is (5.2 ± 0.6) °.

**S3. Significance of gold mirror as a substrate:**

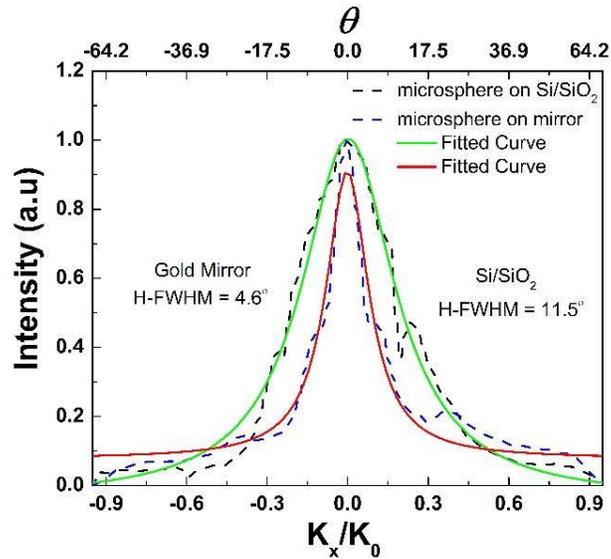

**Figure S3.** Significance of the gold mirror. Dotted curves are intensity profiles of the Fourier plane image shown in Figures 3(b) and 3(c) of the manuscript along the white dotted line. Intensity profile was fitted with a Lorentzian function. The divergence angle for gold mirror and Si/SiO$_2$ substrate is 4.6° and 11.5°, respectively.

**S4. Multiple Fourier plane images for microsphere on Si/SiO$_2$:**

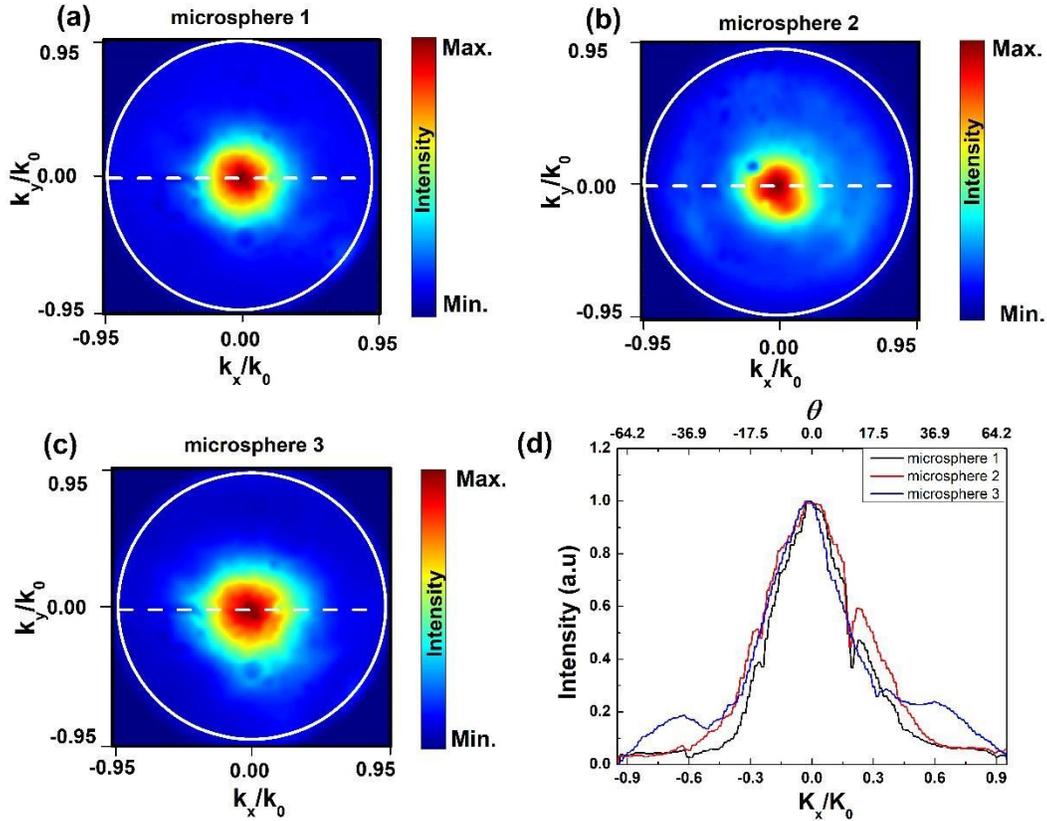

**Figure S4.** Multiple measurements for microspheres on Si/SiO$_2$, (a-c) Fourier plane images of three microspheres on Si/SiO$_2$ substrate, (d) their intensity profile across white dotted line. Divergence angle was calculated by fitting the Lorentzian function for each intensity profile. The average value of divergence angle is (13.2 ± 1.1) °.

**S5. Microsphere size dependence of angular divergence:**

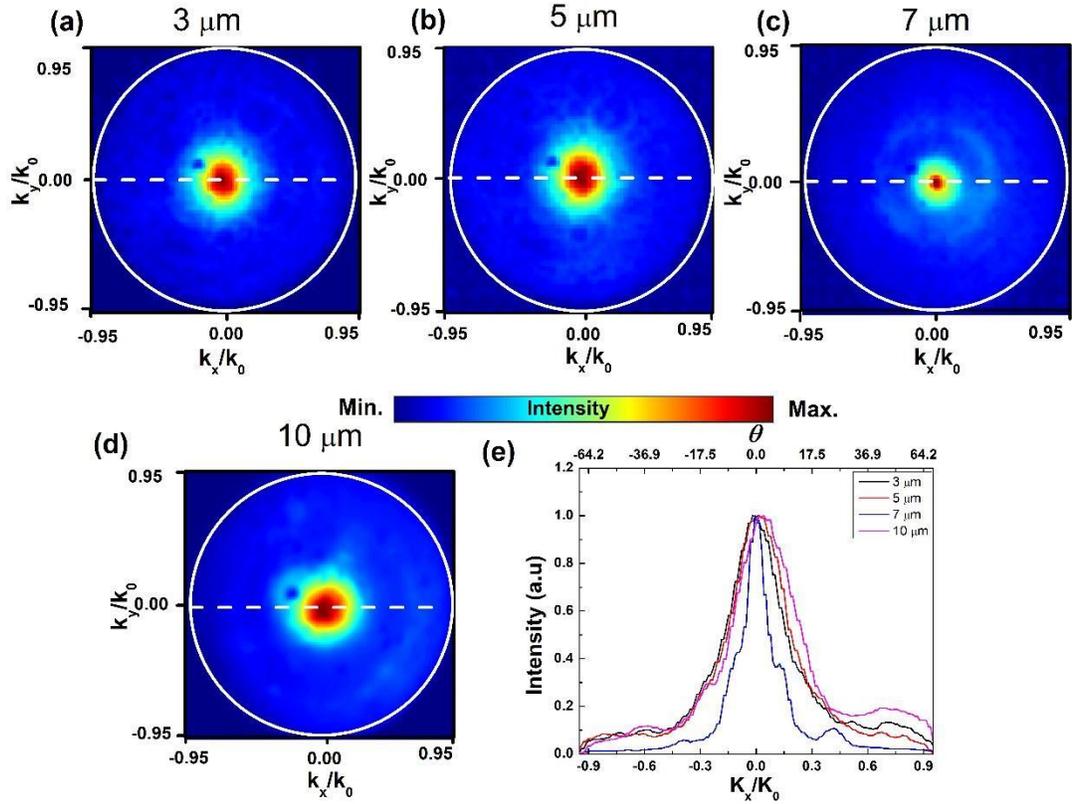

**Figure S5.** Microsphere size dependence, (a-d) Fourier plane image for the microsphere of the size 3, 5, 7, and 10 μm. (e) Intensity profile along the white dotted line in figure (a-d). Divergence angle was calculated by fitting the Lorentzian function for each intensity profile. The divergence angle corresponding to these four Fourier plane images is 9.0°, 9.5, 4.6° and 9.9°, respectively.

**S6. Electric near-field distribution and far field Fourier plane image:**

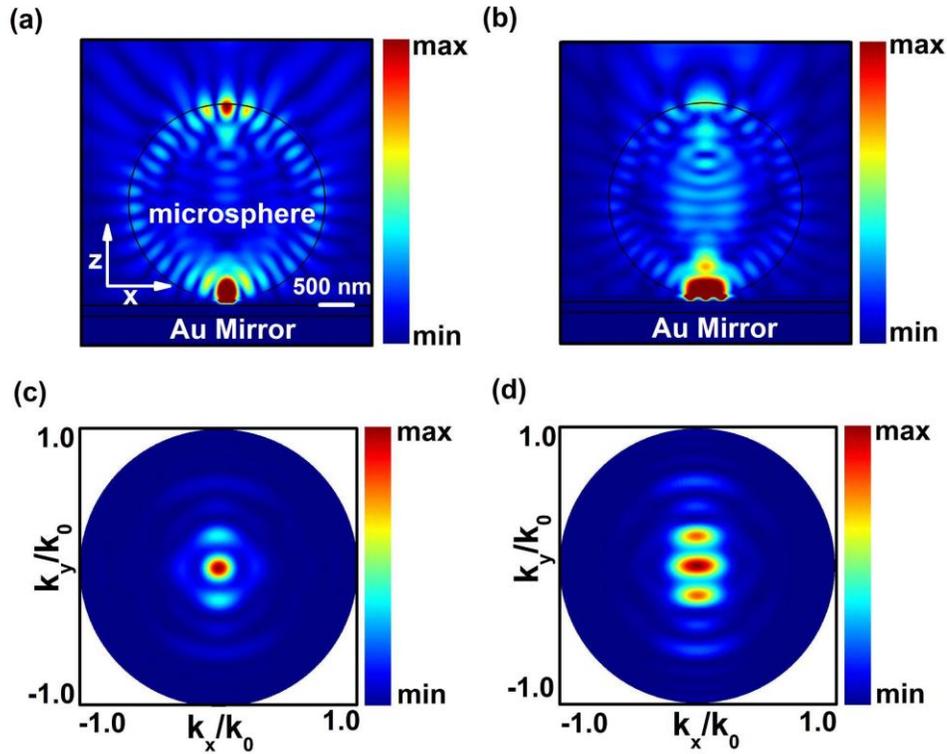

**Figure S6:** Near-field and far-field electric field for a microsphere of diameter 2 μm: Electric near-field distribution when three dipoles were placed (a) 80 nm apart and (b) 240 nm apart, (c) and (d) are the corresponding Fourier plane images.

Defocused excitation results in a bigger spot size which leads to an increase in angular divergence of the emission. Distance between three dipoles were increased to mimic the defocussed excitation. It can be seen from figures S6 (c and d) that the divergence for the smaller distance between dipoles (figure S6 (a) is less. The same model can be used to understand the decrease in the divergence angle in the case of the gold mirror. Since the electric field is more localized in the case of the gold mirror substrate, dipoles were placed closer to mimic the gold mirror case.

**S7. Variation of angular divergence with the numerical aperture of the objective lens**

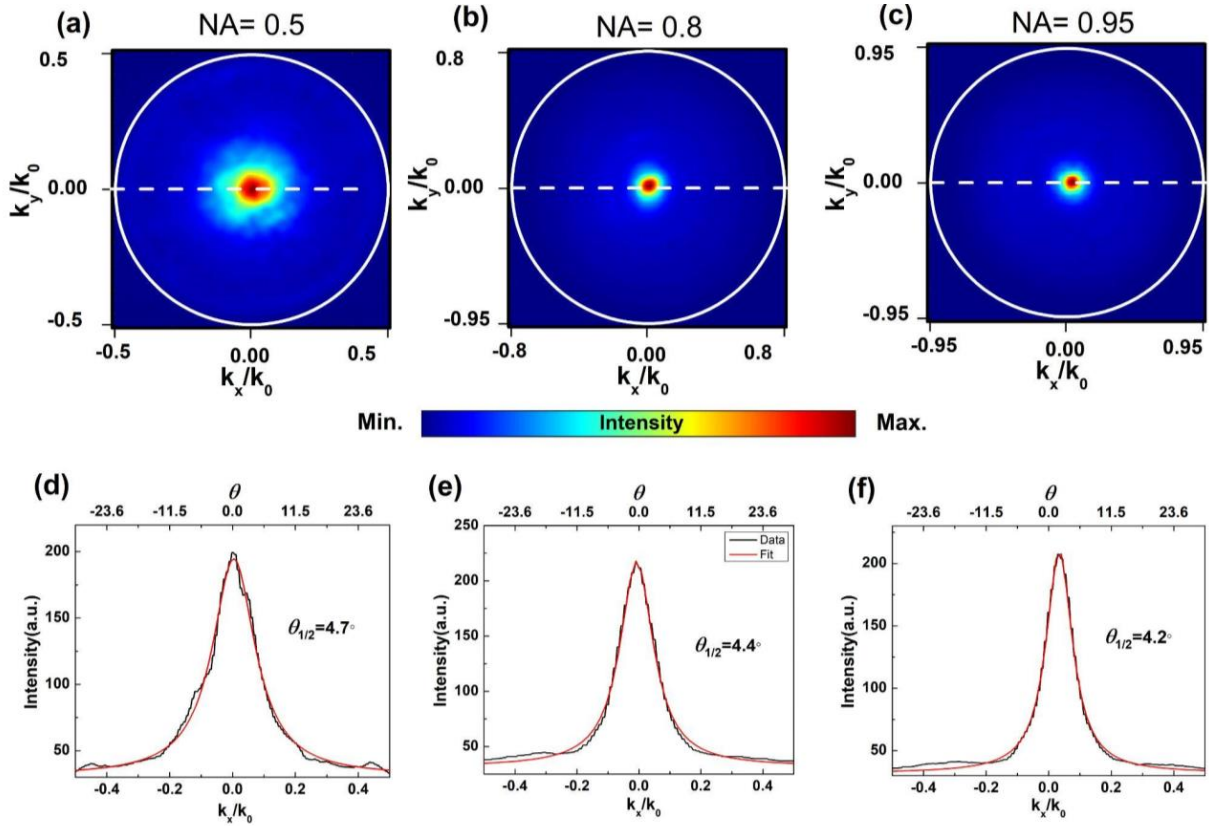

**Figure S7:** Angular divergence variation with the numerical aperture of the objective lens (a-c) are the Fourier plane image corresponding to three different objective lenses with numerical aperture 0.5, 0.8, and 0.95. (d-f) is the intensity profile across the white dotted line in figures (a-c).

No significant changes are observed, in varying the beam waist size. It is expected that the angular divergence will decrease with an increase in the beam waist size. However, this is not observed, because of the high sensitivity of the angular divergence on the beam waist position.

## S8. Electric near-field distribution on glass and gold mirror after focusing the laser beam through microsphere:

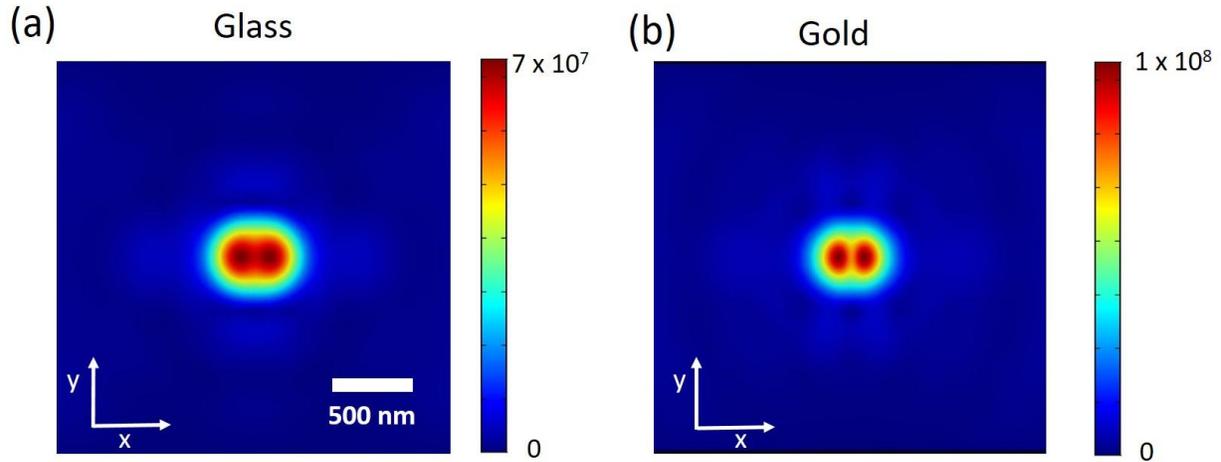

**Figure S8:** Electric near-field distribution in x-y plane for a microsphere of size 3 μm when excited using a Gaussian beam of 532 nm wavelength. (a) When the microsphere is placed on a glass substrate (b) when the microsphere is placed over a gold mirror. In the case of a gold mirror substrate, the electric field is more localized in comparison to a glass substrate. This is because of a hotspot formation around the contact point of the microsphere and the gold mirror. This result implies that for a gold mirror substrate, the emission will be more localized.